\documentclass{elsart3}
\usepackage{graphicx}
\usepackage{amssymb}
\begin{document}
\begin{frontmatter}

\title{Distributed readout detectors using superconducting tunnel junctions}
\author{Iwan Jerjen},
\author{Eugenie Kirk\corauthref{cor}},
\ead{eugenie.kirk@psi.ch}
\author{Elmar Schmid, and Alex Zehnder}
\corauth[cor]{Corresponding author. Tel.: +41-56-310-4351.}

\address{Paul Scherrer Institute, Laboratory for Astrophysics,
5232 Villigen PSI, Switzerland}

\begin{abstract}
Superconducting tunnel junctions (STJs) are able to measure the energy of
single photons in the range from near infrared to X-rays.
They provide simultaneous
information of the impact time and wavelength of an absorbed photon.
The main difficulty of these detectors compared with conventional imaging
detectors lies in their limited pixel number. Each STJ has to be connected
independently and therefore the wiring becomes technologically more
demanding as the number of STJs increases. One approach to solving this
problem is to use a single large absorber and to distribute STJs for
position sensitive signal readout. This configuration is able to detect
single optical photons with an energy resolution close to that of a
single STJ pixel.

We have produced a Ta absorber strip with Ta/Al/AlO$_x$/Al/Nb/Ta junctions
at either end. The energy and position of single photons were measured
simultaneously. The energy resolving power approaches the theoretical limit.
We will present a simple Monte Carlo simulation which reproduces the
measurement exactly.
\end{abstract}

\end{frontmatter}

\section{Introduction}
Superconducting tunnel junctions (STJs) can be used as single photon
detectors with a moderate energy resolution from near infrared to X-rays.
However, a drawback of this technology is
the limited number and size of the junctions available in one detector.

One approach to this problem is to separate the absorption and the
read-out processes by the use of one large absorber and several
distributed junctions for readout \cite{Wilson,Verhoeve,Kirk}.
The sum of the signals of all
junctions measures the energy of the photon and the difference of
the signal amplitudes allows to calculate the position of the photon impact.

In this paper we will present the response of a strip detector to optical
photons and compare it with a Monte Carlo simulation.

\section{Experiment}
We deposited a $135\, \mu \mathrm{m}$ long, $31.5\, \mu \mathrm{m}$
wide and $100\, \mathrm{nm}$ thick Ta absorber layer ($RRR \approx 25$)
on a sapphire substrate.
At each end, on top of the absorber, we fabricated
$25 \times 25\, \mu \mathrm{m}^2$  Ta-Al junctions with $38\, \mathrm{nm}$
thick Al layers (including an intermediate thin Nb seed layer).
The device was cooled down to a temperature of
$0.32\, \mathrm{K}$. We biased each junction independently at about
$100\, \mu \mathrm{V}$ where the thermal current was $\sim 200\, \mathrm{pA}$.
A pulsed $592\, \mathrm{nm}$ LED ($17\, \mathrm{nm}$ FWHM)
served as light source.
We read out the signals with a charge-voltage-conversion amplifier.
The signals were digitalized and stored in a file for offline analysis.

\begin{figure}[h]
   \centering
   \includegraphics[width=0.90\linewidth]{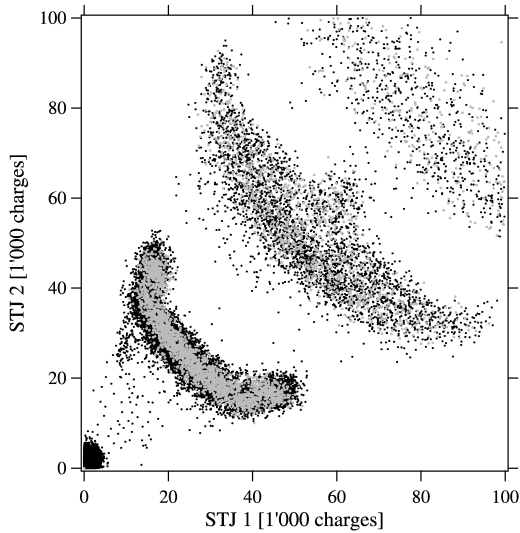}
   \caption{The response of a $135\, \mu \mathrm{m}$ Ta strip to
   $592\, \mathrm{nm}$ photons is shown. The signals of the two STJs
   at the end of the strip are plotted against each other.
   The black dots are measured values and the grey points are simulated.
   Electronic noise as well as single, double and
   triple photon events are visibly separated
   (diagonally from bottom left to top right).
    }
   \label{fig1.f}
\end{figure}

Figure~\ref{fig1.f} shows the response (black dots) of the two junctions
to the absorption of $592\, \mathrm{nm}$ photons in the Ta layer.
The black spot near zero is noise.
The first and second banana curves correspond to the absorption of
one and two photons, respectively.
The small curving of the middle part means that
the quasiparticles diffuse very fast (compared to the loss processes)
across the entire strip.
The increase of the signal amplitudes at the end of the strip is
due to a degradation of the gap under the junctions which results
in a higher number of initially created quasiparticles \cite{Verhoeve}.

\begin{figure}[h]
   \centering
   \includegraphics[width=1.0\linewidth]{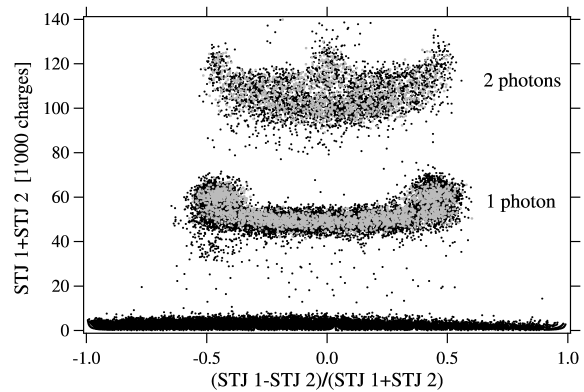}
   \caption{The sum of the signals is plotted against the normalized
   difference which contains the position information. The black spots
   are the measured values, the grey points are simulated.
    }
   \label{fig2.f}
\end{figure}

Figure~\ref{fig2.f} shows the same measurement by displaying the sum
of the signals (energy) as a function of the normalized difference (position).
The histogram of the center values of the single photon events
has a FWHM of 6460 charges.
An electronic noise of 4490 charges was measured by applying a test pulse.
The width of the light source amounts to 1440 charges.
By subtracting these external noise sources from the total noise
we obtain an intrinsic noise of 4420 charges which corresponds to a
$0.19\, \mathrm{eV}$ resolution or a resolving power of 11.
The theoretical resolving power obtained by taking into account the
Fano factor and the tunnel noise only and by assuming that all the
created quasiparticles take part in the tunnel process is 16 \cite{Verhoeve}.
However, this theoretical value can not be reached since a significant
fraction of the quasiparticles is lost
before contributing to the signal by tunneling.

\section{Monte Carlo simulation}
We propose a simple two-dimensional Monte Carlo simulation for
modeling our strip detector. The absorption of a photon will
create $N_0 = E_{\gamma} / 1.7 \Delta$ quasiparticles,
where $E_{\gamma}$
is the photon energy and $\Delta$ the Ta gap energy.
To this number $N_0$ we added a Gaussian noise corresponding
to the bandwidth of the light source and the Fano noise.
At each simulation step every single quasiparticle moves a distance
$d$ in an arbitrary direction. The distance $d$ is a model parameter
chosen to be small compared to the junction dimensions.
If a quasiparticle would move out of the Ta strip it is set back
to the borders of the layer (i.e.\ it is not reflected and will move
in an arbitrary direction again at the next simulation step).
During each step there is a certain probability that a quasiparticle
will be lost ($P_{\mathrm{loss,Abs}}$ and $P_{\mathrm{loss,STJ}}$) or,
if it stays within the junction borders, that it is trapped ($P_{\mathrm{trap}}$).
There are two loss probabilities to take into account that the quasiparticle
lifetime in the STJ is higher than in the absorber area because in the
junction area one has to consider the mean lifetime of quasiparticles
in Ta and Al \cite{Jerjen1}.
Once a quasiparticle is trapped it can't move out again in our model.
Finally there is a certain probability that a quasiparticle staying
within the junction borders contributes to the signal ($P_{\mathrm{sig}}$)
whereupon it is taken out of the simulation.
To the number of read out charges the tunnel noise is added to obtain
the final signal. Since the Ta gap is slightly reduced in the junction
area we introduced also a parameter $F < 1$ taking into account that
the number of charges created outside the junctions is smaller.

Agreement between experimental data and Monte Carlo simulations
(black and grey points, respectively,
as shown in Figs.~\ref{fig1.f} and \ref{fig2.f}) was obtained by empirically
tuning the simulation parameters, yielding the best fit parameters:
$d = 6\, \mu \mathrm{m}$,  $P_{\mathrm{loss,Abs}} = 7 \cdot 10^{-4}$
and $P_{\mathrm{loss,STJ}} = 5.1 \cdot 10^{-4}$,
$P_{\mathrm{trap}} = 3.0 \cdot 10^{-3}$,
$P_{\mathrm{sig}} = 2.0 \cdot 10^{-3}$,
$F = 0.875$.
Because the loss probabilities are calculated per simulation step
for a given quasiparticle propagation length $d$, one could, in principle,
determine the corresponding loss rates and the diffusion speed via
diffusion model if one of those values were known.

\section{Summary}
Our measurements confirm the results of Ref.~\cite{Verhoeve}, proving
that a good energy resolution can be achieved with a distributed
readout scheme. We have also shown that a simple Monte Carlo simulation
reproduces the experimental values.
From the simulation parameter $F = 0.875$ we conclude
that the relevant energy gap for
absorption in the Ta layer under
the junction is about 12.5\% smaller than in the strip.
This energy difference of
$88\, \mu \mathrm{eV}$ is on the order of the thermal energy
($k_B T = 28\, \mu \mathrm{eV}$) and thus the quasiparticles
are not expected to be trapped totally in the junction area.
This agrees with the fact that the photons absorbed at one end
of the strip are also detected by the junction at the opposite end.
Furthermore, this indicates that the
mean energy of the quasiparticles which are created after
absorption of a photon in the Ta layer under the junction is higher
than the gap energy at
the barrier as measured by the $IV$ curve which is
about $450\, \mu \mathrm{eV}$.
This agrees with the 4-quasiparticle-populations model presented in
our former work \cite{Jerjen1,Jerjen2}.

Combining that model with a three-dimensional
Monte Carlo simulation would allow us to calculate and,
by variation of simulation parameters, to optimize the responsivity and
energy resolution of STJ distributed readout detectors.
As an example, we found that a moderately thicker Al trapping layer
is favourable.
On the other hand, the less significant contribution of the absorber
quality to the device performance as found in our experiments
could be verified.

\section*{Acknowledgements}
We are grateful to M.\ Furlan and Ph.\ Lerch for valuable discussions
and to F.\ Burri for technical support.

\end{document}